\RequirePackage{lineno}
\documentclass[aps,prl,preprint,tightenlines,superscriptaddress,showpacs,byrevtex]{revtex4-1}

\usepackage{graphicx} 
\usepackage{dcolumn}  
\usepackage{color} %
\RequirePackage{xspace}
\usepackage{relsize}

\graphicspath{{ps}}

\def\Bbar     {\kern 0.2em\overline{\kern -0.2em B}{}\xspace}
\def\Dbar     {\kern 0.2em\overline{\kern -0.2em D}{}\xspace}
\def\to       {\ensuremath{\rightarrow}\xspace} 
\def\bzetapiz {\ensuremath{B^0\to\eta \pi^0}\xspace} 
\def\qbar     {\kern 0.2em\overline{\kern -0.2em q}{}\xspace}
\def\evtgen   {\mbox{\textsc{EvtGen}}\xspace}
\def\geant    {\mbox{\textsc{Geant3}}\xspace}
\def\photos   {\mbox{\textsc{Photos}}\xspace}
\def\infb     {fb$^{-1}$}

\def\gevm     {GeV/$c^2$}

\begin{document}
\vspace*{-2\baselineskip}
\resizebox{!}{3cm}{\includegraphics{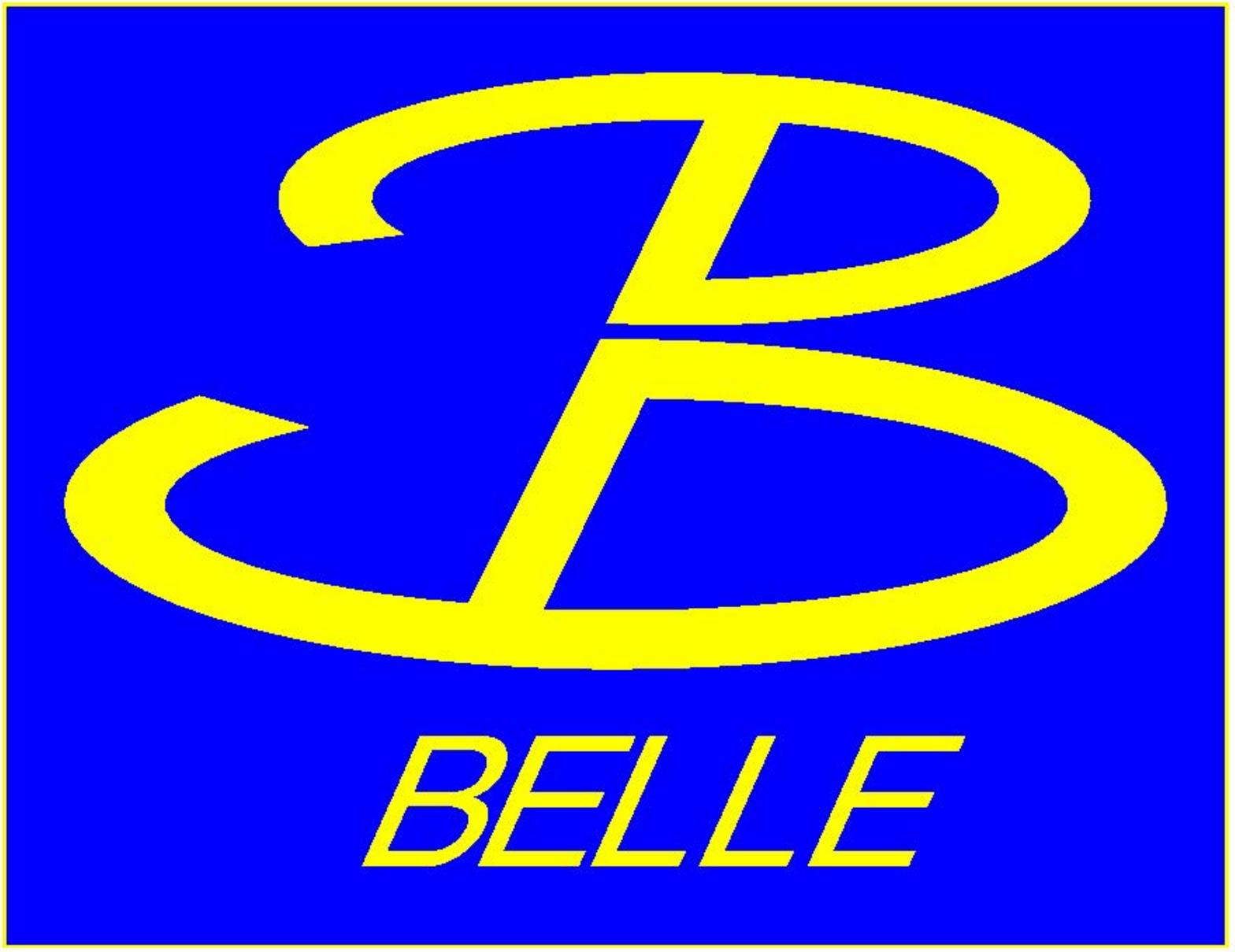}}

\preprint{\vbox{ \hbox{   }
			\hbox{\textbf  {Belle Preprint  2015-2}}
                             \hbox{\textbf {KEK Preprint  2014-45}}
                            \hbox{\textbf  {UCHEP Preprint  2015-2}}
}}

\vskip -1.5cm
\title{ \quad\\[1.0cm] Evidence for the decay \mbox{\boldmath$B^0\rightarrow\eta\pi^0$} }

\noaffiliation
\affiliation{University of the Basque Country UPV/EHU, 48080 Bilbao}
\affiliation{Beihang University, Beijing 100191}
\affiliation{University of Bonn, 53115 Bonn}
\affiliation{Budker Institute of Nuclear Physics SB RAS and Novosibirsk State University, Novosibirsk 630090}
\affiliation{Faculty of Mathematics and Physics, Charles University, 121 16 Prague}
\affiliation{University of Cincinnati, Cincinnati, Ohio 45221}
\affiliation{Deutsches Elektronen--Synchrotron, 22607 Hamburg}
\affiliation{Justus-Liebig-Universit\"at Gie\ss{}en, 35392 Gie\ss{}en}
\affiliation{Gifu University, Gifu 501-1193}
\affiliation{SOKENDAI (The Graduate University for Advanced Studies), Hayama 240-0193}
\affiliation{Hanyang University, Seoul 133-791}
\affiliation{University of Hawaii, Honolulu, Hawaii 96822}
\affiliation{High Energy Accelerator Research Organization (KEK), Tsukuba 305-0801}
\affiliation{IKERBASQUE, Basque Foundation for Science, 48013 Bilbao}
\affiliation{Indian Institute of Technology Guwahati, Assam 781039}
\affiliation{Indian Institute of Technology Madras, Chennai 600036}
\affiliation{Indiana University, Bloomington, Indiana 47408}
\affiliation{Institute of High Energy Physics, Chinese Academy of Sciences, Beijing 100049}
\affiliation{Institute of High Energy Physics, Vienna 1050}
\affiliation{Institute for High Energy Physics, Protvino 142281}
\affiliation{INFN - Sezione di Torino, 10125 Torino}
\affiliation{Institute for Theoretical and Experimental Physics, Moscow 117218}
\affiliation{J. Stefan Institute, 1000 Ljubljana}
\affiliation{Kanagawa University, Yokohama 221-8686}
\affiliation{Institut f\"ur Experimentelle Kernphysik, Karlsruher Institut f\"ur Technologie, 76131 Karlsruhe}
\affiliation{King Abdulaziz City for Science and Technology, Riyadh 11442}
\affiliation{Department of Physics, Faculty of Science, King Abdulaziz University, Jeddah 21589}
\affiliation{Korea Institute of Science and Technology Information, Daejeon 305-806}
\affiliation{Korea University, Seoul 136-713}
\affiliation{Kyungpook National University, Daegu 702-701}
\affiliation{\'Ecole Polytechnique F\'ed\'erale de Lausanne (EPFL), Lausanne 1015}
\affiliation{Faculty of Mathematics and Physics, University of Ljubljana, 1000 Ljubljana}
\affiliation{Luther College, Decorah, Iowa 52101}
\affiliation{University of Maribor, 2000 Maribor}
\affiliation{Max-Planck-Institut f\"ur Physik, 80805 M\"unchen}
\affiliation{School of Physics, University of Melbourne, Victoria 3010}
\affiliation{Moscow Physical Engineering Institute, Moscow 115409}
\affiliation{Moscow Institute of Physics and Technology, Moscow Region 141700}
\affiliation{Graduate School of Science, Nagoya University, Nagoya 464-8602}
\affiliation{Kobayashi-Maskawa Institute, Nagoya University, Nagoya 464-8602}
\affiliation{Nara Women's University, Nara 630-8506}
\affiliation{National Central University, Chung-li 32054}
\affiliation{National United University, Miao Li 36003}
\affiliation{Department of Physics, National Taiwan University, Taipei 10617}
\affiliation{H. Niewodniczanski Institute of Nuclear Physics, Krakow 31-342}
\affiliation{Niigata University, Niigata 950-2181}
\affiliation{University of Nova Gorica, 5000 Nova Gorica}
\affiliation{Osaka City University, Osaka 558-8585}
\affiliation{Pacific Northwest National Laboratory, Richland, Washington 99352}
\affiliation{Peking University, Beijing 100871}
\affiliation{University of Science and Technology of China, Hefei 230026}
\affiliation{Seoul National University, Seoul 151-742}
\affiliation{Soongsil University, Seoul 156-743}
\affiliation{Sungkyunkwan University, Suwon 440-746}
\affiliation{School of Physics, University of Sydney, NSW 2006}
\affiliation{Department of Physics, Faculty of Science, University of Tabuk, Tabuk 71451}
\affiliation{Tata Institute of Fundamental Research, Mumbai 400005}
\affiliation{Excellence Cluster Universe, Technische Universit\"at M\"unchen, 85748 Garching}
\affiliation{Toho University, Funabashi 274-8510}
\affiliation{Tohoku University, Sendai 980-8578}
\affiliation{Department of Physics, University of Tokyo, Tokyo 113-0033}
\affiliation{Tokyo Institute of Technology, Tokyo 152-8550}
\affiliation{Tokyo Metropolitan University, Tokyo 192-0397}
\affiliation{University of Torino, 10124 Torino}
\affiliation{CNP, Virginia Polytechnic Institute and State University, Blacksburg, Virginia 24061}
\affiliation{Wayne State University, Detroit, Michigan 48202}
\affiliation{Yamagata University, Yamagata 990-8560}
\affiliation{Yonsei University, Seoul 120-749}
  \author{B.~Pal}\affiliation{University of Cincinnati, Cincinnati, Ohio 45221} 
  \author{A.~J.~Schwartz}\affiliation{University of Cincinnati, Cincinnati, Ohio 45221} 
  \author{A.~Abdesselam}\affiliation{Department of Physics, Faculty of Science, University of Tabuk, Tabuk 71451} 
  \author{I.~Adachi}\affiliation{High Energy Accelerator Research Organization (KEK), Tsukuba 305-0801}\affiliation{SOKENDAI (The Graduate University for Advanced Studies), Hayama 240-0193} 
  \author{H.~Aihara}\affiliation{Department of Physics, University of Tokyo, Tokyo 113-0033} 
  \author{S.~Al~Said}\affiliation{Department of Physics, Faculty of Science, University of Tabuk, Tabuk 71451}\affiliation{Department of Physics, Faculty of Science, King Abdulaziz University, Jeddah 21589} 
  \author{K.~Arinstein}\affiliation{Budker Institute of Nuclear Physics SB RAS and Novosibirsk State University, Novosibirsk 630090} 
  \author{D.~M.~Asner}\affiliation{Pacific Northwest National Laboratory, Richland, Washington 99352} 
  \author{V.~Aulchenko}\affiliation{Budker Institute of Nuclear Physics SB RAS and Novosibirsk State University, Novosibirsk 630090} 
 \author{T.~Aushev}\affiliation{Moscow Institute of Physics and Technology, Moscow Region 141700}\affiliation{Institute for Theoretical and Experimental Physics, Moscow 117218} 
% ...
\collaboration{The Belle Collaboration}

\begin{abstract}
We report a search for the charmless hadronic decay 
\bzetapiz\ with a data sample corresponding to an
integrated luminosity of 694 $\rm fb^{-1}$ containing
$753\times10^6$ $B\Bbar$ pairs. The data were collected
by the Belle experiment running on the  $\Upsilon(4S)$
resonance at the KEKB $e^+e^-$ collider. We measure a
branching fraction 
$\mathcal{B}(B^0\to\eta\pi^0)=(4.1^{+1.7+0.5}_{-1.5-0.7})\times 10^{-7}$,
where the first uncertainty  is statistical and the second is systematic. Our measurement gives an upper limit of $\mathcal{B}(B^0\to\eta\pi^0)<6.5\times 10^{-7}$  at 90\% confidence level.
The signal has a significance of $3.0$ standard deviations and constitutes 
the first evidence for this decay mode. 
\end{abstract}


\maketitle

\tighten

\begin{figure}[htb]
\includegraphics[width=0.35\textwidth]{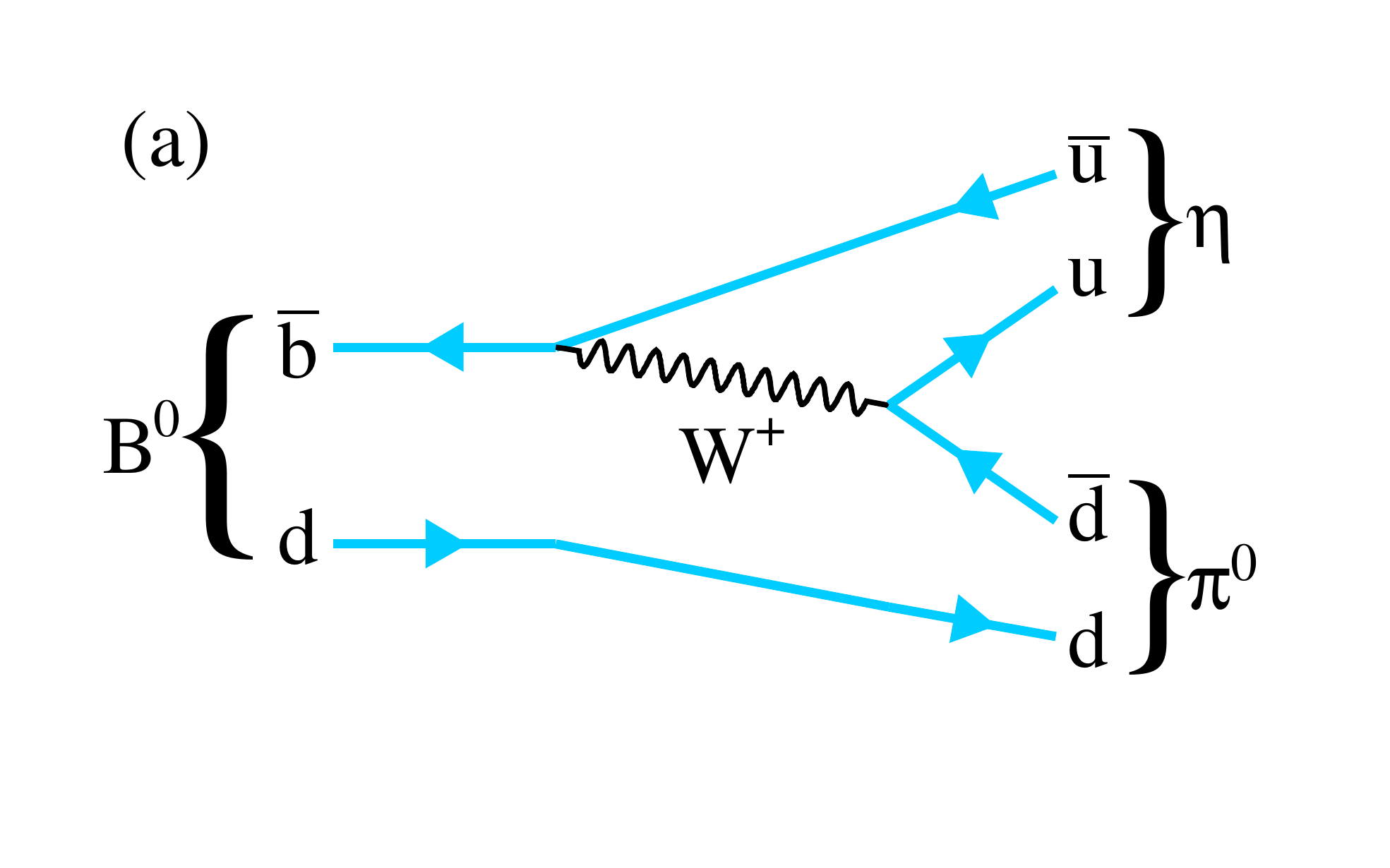}%
\includegraphics[width=0.35\textwidth]{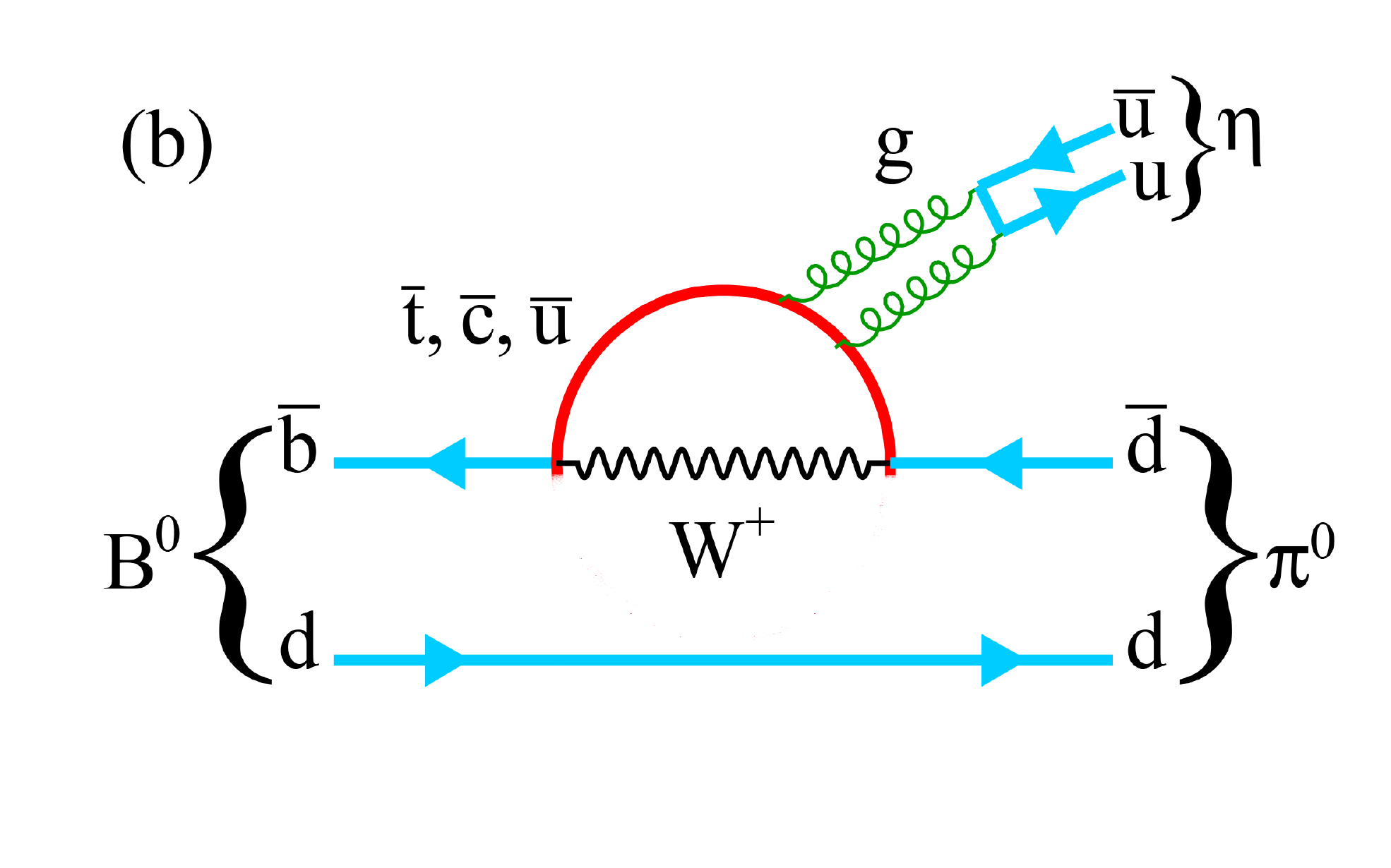}%
\vskip -0.5cm
\caption{\small (a) Tree  and (b) penguin diagram 
contributions to \bzetapiz. }
\label{fig:feynman}
\end{figure}

Two-body charmless hadronic decays of $B$ mesons are
important for determining Standard Model  parameters
and for detecting the presence of new physics~\cite{newphysics}. The
decay \bzetapiz\ proceeds mainly via a $b\to u$ Cabibbo- and
color-suppressed ``tree'' diagram, and via a $b\to d$ 
``penguin" diagram~\cite{charge-conjugate}, as shown
in Fig.~\ref{fig:feynman}. 
The branching fraction can
be used to constrain isospin-breaking effects on the value of
$\sin2\phi_2~(\sin 2\alpha)$  measured in $B\to\pi\pi$
decays~\cite{Gronau:2005pq,Gardner:2005pq}.
It can also be used
to constrain  $CP$-violating parameters ($C^{}_{\eta' K}$
and $S^{}_{\eta' K}$) governing the time dependence of
$B^0\to\eta' K^0$ decays~\cite{etapKs_bound}.
The branching fraction is estimated 
using QCD factorization~\cite{qcd}, soft collinear effective
field theory~\cite{Williamson:2006hb}, and flavor SU(3) symmetry~\cite{su3}
and is found to be in the range $(2 - 12)\times10^{-7}$.

Several experiments~\cite{Albrecht:1990am, Acciarri:1995bx, Richichi:1999kj, Chang:2004fz, Aubert:2008fu}, including Belle, have searched for this decay mode.
The current most stringent limit on the branching fraction
is $\mathcal{B}(B^0\to\eta\pi)^0<1.5\times10^{-6}$ at 
90\% confidence level (C.L.)~\cite{Aubert:2008fu}. Here we update our
previous result~\cite{Chang:2004fz}
using the full data set of the Belle experiment running on
the $\Upsilon(4S)$ resonance at the KEKB asymmetric-energy 
$e^+e^-$ collider~\cite{KEKB}. This data set corresponds to
$753\times10^{6}$ $B\Bbar$ pairs, which is a factor of 5
larger than that used previously. The analysis presented
here also uses improved tracking, photon reconstruction,
and continuum suppression algorithms.

The Belle detector is a large-solid-angle magnetic spectrometer consisting
of a  silicon vertex detector (SVD), a 50-layer central
drift chamber (CDC), an array of aerogel threshold Cherenkov counters (ACC),
a barrel-like arrangement of time-of-flight scintillation counters (TOF),
and an electromagnetic calorimeter (ECL) comprising  CsI(Tl) crystals. 
These detector components are 
located inside a superconducting solenoid coil that provides a 1.5~T
magnetic field.  An iron flux-return located outside the coil (KLM)
is instrumented to detect $K_L^0$ mesons and to identify muons.
Two inner detector configurations were used: a 2.0 cm beampipe
and a three-layer SVD were used for the first 
123~\infb\ of data, 
while a 1.5 cm beampipe, a four-layer SVD, and a 
small-cell inner drift chamber were used for the remaining
571~\infb\ of data.
The detector is described in detail elsewhere~\cite{Belle,svd2}.

An  ECL cluster not matched to any track  is identified as a photon candidate.
The timing of the energy deposited in the ECL must be  consistent 
with the beam collision time identified at the trigger level. All photon candidates
are required to have an energy greater than 50 MeV.
We reconstruct $\pi^0\to\gamma\gamma$ decays by pairing
together photon candidates and requiring that the
$\gamma\gamma$ invariant mass be in the range 0.115-0.155~\gevm. 
This corresponds to $\pm3.5\sigma$ around  the nominal  $\pi^0$ mass~\cite{PDG}.
Photon candidates in the end cap regions are required to have an 
energy greater than 100 MeV for $\pi^0$ reconstruction.
To improve the $\pi^0$ momentum  resolution, we perform
a mass-constrained fit and require that the resulting $\chi^2$
be less than 50.

Candidate $\eta$ mesons are reconstructed
via $\eta\to\gamma\gamma$ ($\eta_{\gamma\gamma}$)
and $\eta\to\pi^+\pi^-\pi^0$ ($\eta_{3\pi}$) decays.
At least one of the photons in an $\eta_{\gamma\gamma}$ 
candidate must have an energy greater than 100 MeV.
To reduce combinatorial background
from low-energy photons, $\eta_{\gamma\gamma}$ candidates
are required to satisfy $|E_1-E_2|/(E_1+E_2)<0.9$, where
$E_1$ and $ E_2$ are the two  photon energies. 
Photons used for $\eta_{\gamma\gamma}$
reconstruction are not allowed to pair with any other photon
 to form a $\pi^0$ candidate. 
Candidate $\eta_{3\pi}$ mesons are reconstructed by combining a
$\pi^0$ with a pair of oppositely charged tracks. These tracks
are required to have a distance of closest approach with respect
to the interaction point  along the $z$ axis  (antiparallel to  the $e^+$ beam) of
$|dz|<3.0$~cm, and in the transverse plane of $|dr|<0.3$  cm. 
Pions are identified using information obtained
from the CDC ($dE/dx$), the TOF, and the ACC. This information is combined
to form a likelihood ($\mathcal{L}$) for hadron identification.
We require that charged tracks satisfy
$\mathcal{L}_K/(\mathcal{L}_{\pi}+\mathcal{L}_K)<0.4$, where
$\mathcal{L}_K\,(\mathcal{L}_{\pi})$ is the likelihood of the
track being a kaon (pion). We reject tracks whose response in
the ECL and KLM are consistent with that of an electron or muon. 
The pion identification efficiency is 91.2\% and the probability for a kaon to be 
 misidentified as  a pion is 5.4\%.
We require that the invariant mass of $\eta_{\gamma\gamma}$
and $\eta_{3\pi}$ candidates be in the ranges 0.500-0.575~\gevm\ 
and 0.538-0.557~\gevm, respectively,  corresponding to $\pm2.5\sigma$ and $\pm3.0\sigma$ in resolution  around the nominal $\eta$ mass~\cite{PDG}.
For selected $\eta$ candidates, a mass-constrained fit is
performed to improve the momentum resolution.

Candidate $B$ mesons are identified using the beam-energy-constrained
mass, $M_{\rm bc}=\sqrt{E_{\rm beam}^2-|\vec{p}^{}_B|^2 c^2 }/c^2$, and the energy
difference, $\Delta E=E_B-E_{\rm beam}$, where $E_{\rm beam}$ is the beam
energy, and $E_B$ and $\vec{p}^{}_B$ are the energy and momentum, respectively,
of the $B$ candidate. All quantities are evaluated in the $\Upsilon(4S)$
center-of-mass (CM) frame. We require that events satisfy
$M_{\rm bc} > 5.24$~\gevm\ and $-0.30~{\rm GeV} < \Delta E< 0.25~{\rm GeV}$.
We calculate signal yields in a smaller  region 
$M_{\rm bc} > 5.27$~\gevm\ and $-0.21~{\rm GeV} < \Delta E< 0.15~{\rm GeV}$.

Charmless hadronic decays suffer from large backgrounds arising
from continuum $e^+e^-\to q\qbar~(q=u,d,s,c)$ production.  
To suppress this background, we use a multivariate analyzer based on 
a neural network (NN)~\cite{Feindt:2006pm}. 
The NN uses the event topology and $B$-flavor tagging
information~\cite{TaggingNIM} to discriminate continuum 
events, which tend to be jetlike, from spherical $B\Bbar$ events.
The event shape variables include
16 modified Fox-Wolfram moments~\cite{SFW}, the cosine of the
angle between the $z$ axis and the $B$ flight direction, and the
cosine of the angle between the thrust axis~\cite{Brandt:1964sa}
of the $B$ candidate and the thrust axis of the rest of the event.
All quantities are evaluated in the $\Upsilon(4S)$ CM frame.

The NN technique requires a training procedure and, after training,
achieves excellent separation between signal and background from
the output variable $C_{NB}$. This variable ranges from $-1$ to $+1$: a value
closer to $-1$  ($+1$) is more likely to identify  a background (signal) event. 
The training samples used consist
of Monte Carlo (MC) $\bzetapiz$ events for signal and an
$M_{\rm bc}$--$\Delta E$ sideband from data for background. The regions of $5.200~{\rm GeV}/c^2 < M_{\rm bc} <5.265~{\rm GeV}/c^2$, $\Delta E<-0.23~{\rm GeV}$ and  $\Delta E > 0.17~{\rm GeV}$ define the $M_{\rm bc}$--$\Delta E$ sideband.
Independent samples are used to test the NN performance.
The MC samples are obtained using \evtgen~\cite{Lange:2001uf}
for event generation and the \geant~\cite{geant3} package
for modeling of the detector response. 
Final-state radiation is taken into account using the
\photos~\cite{Golonka:2005pn} package.

We require $C_{NB}>-0.1$, which rejects
approximately 85\% of continuum background events while retaining
90\% of signal events. We subsequently translate 
$C_{NB}$ to $C'_{NB}$, defined as
\begin{linenomath}
\begin{equation}
C'_{NB} =\ln\left(\frac{C_{NB}-C^{\rm min}_{NB}}{C^{\rm max}_{NB}-C_{NB}}\right)
\end{equation} 
\end{linenomath}
where $C^{\rm min}_{NB}= -0.1$ and $C_{NB}^{\rm max}$ is
the maximum value of $C_{NB}$ obtained from a large sample of signal MC decays. 
This translation is advantageous as the $C'_{NB}$ distribution
for both signal and background is well described by a  sum of
Gaussian functions. 

After applying all  selection criteria, 
2\% (7\%)  of events have more than one $B^0\to\eta_{\gamma\gamma}\pi^0$ ($B^0\to\eta_{3\pi}\pi^0$) candidate.
For these events, we retain the $\bzetapiz$ candidate 
 with the smallest $\chi^2$ value resulting from the  $\eta$ or, if necessary, $\pi^0$  mass-constrained fits.  According to MC simulations, this criterion chooses the correct
$B$ candidate 63\% (77\%) of the time for 
$B^0\to\eta_{\gamma\gamma}\pi^0$ ($B^0\to\eta_{3\pi}\pi^0$).

We calculate signal yields using an unbinned extended maximum
likelihood fit to the variables $M_{\rm bc}$, $\Delta E$, and
$C'_{NB}$. The likelihood function is defined as
\begin{linenomath}
\begin{equation}
\mathcal{L} = e^{-\sum_j Y_j}\cdot 
\prod_i^N \left( \sum_j Y_j \mathcal{P}_j(M_{\rm bc}^i, \Delta E^i, C'^i_{NB} )\right),
\end{equation} 
\end{linenomath}
where $N$ is the total number of events,
$\mathcal{P}_j(M_{\rm bc}^i, \Delta E^i, C'^i_{NB})$ is the
probability density function (PDF) of signal or background component
$j$ for event $i$, and $j$ runs over all signal and background
components. $Y_j$ is the yield of component~$j$.
The background components consist of  $e^+e^-\to q\bar{q}$ continuum
events, generic $b\to c$ processes, and charmless rare processes. 
The latter two backgrounds are small compared to the $q\qbar$
continuum events and are studied using MC simulations. We find
that no $b\to c$  events pass our selection criteria.
The charmless rare background, however, shows peaking structure in
the $M_{\rm bc}$ distribution, most of which arises from
$B^+\to\eta\rho^+$ decays. 

Correlations among the fit variables are found to be small, and 
thus we factorize the PDFs as
\begin{linenomath}
\begin{equation}
\mathcal{P}_j(M_{\rm bc}, \Delta E, C'_{NB}) = 
\mathcal{P}_j(M_{\rm bc})\cdot\mathcal{P}_j(\Delta E)\cdot\mathcal{P}_j(C'_{NB}).
\end{equation}
\end{linenomath}
All PDFs for $C'_{NB}$ are modeled with the sum of two Gaussian functions.
The $M_{\rm bc}$ and $\Delta E$ PDFs for signal events are modeled
with ``crystal ball'' (CB) functions~\cite{crystalball}.
The peak positions and resolutions 
in the signal $M_{\rm bc}$, $\Delta E$, and $C'_{NB}$
are adjusted according to data-MC differences observed
in a high statistics control sample of $B^0\to\Dbar^0(\to K^+\pi^-\pi^0)\pi^0$
decays.
This decay has four photons, as do signal decays, and its 
topology is identical to that of $B^0\to\eta^{}_{3\pi}\pi^0$. 
The $C'_{NB}$  PDF of the continuum background is also adjusted by comparing  data and continuum  MC samples  in the $M_{\rm bc}$  sideband (5.200-5.265~\gevm).
The $\Delta E$ PDF for continuum background is modeled with
a second-order polynomial, while the $M_{\rm bc}$ PDF is modeled
with an ARGUS function~\cite{Albrecht:1990am}. 
The $M_{\rm bc}$ and $\Delta E$ PDFs for charmless rare background are
modeled with one-dimensional nonparametric PDFs based on kernel
estimation~\cite{Cranmer:2000du}.
In addition to the fitted yields $Y^{}_j$, the $M_{\rm bc}$ and 
$\Delta E$ PDF parameters for continuum background are also
floated, except for the end point of the ARGUS function.
All other parameters are fixed to the corresponding MC values.
To test the stability of the fitting procedure, numerous fits
are performed to large ensembles of MC events.

\begin{figure}[h!t!p!]
\begin{center}
    \includegraphics[width=0.35\textwidth]{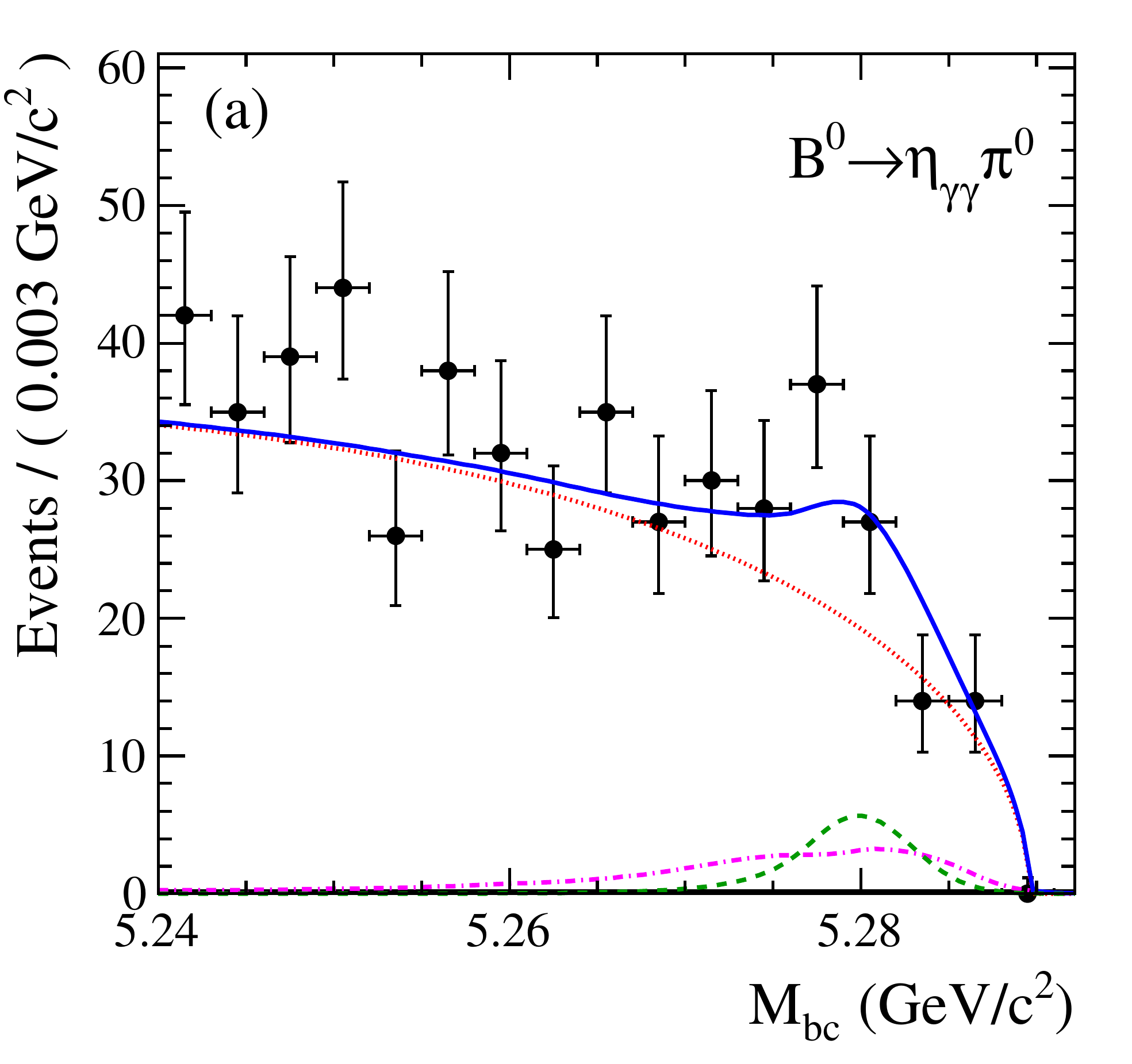}%
    \includegraphics[width=0.35\textwidth]{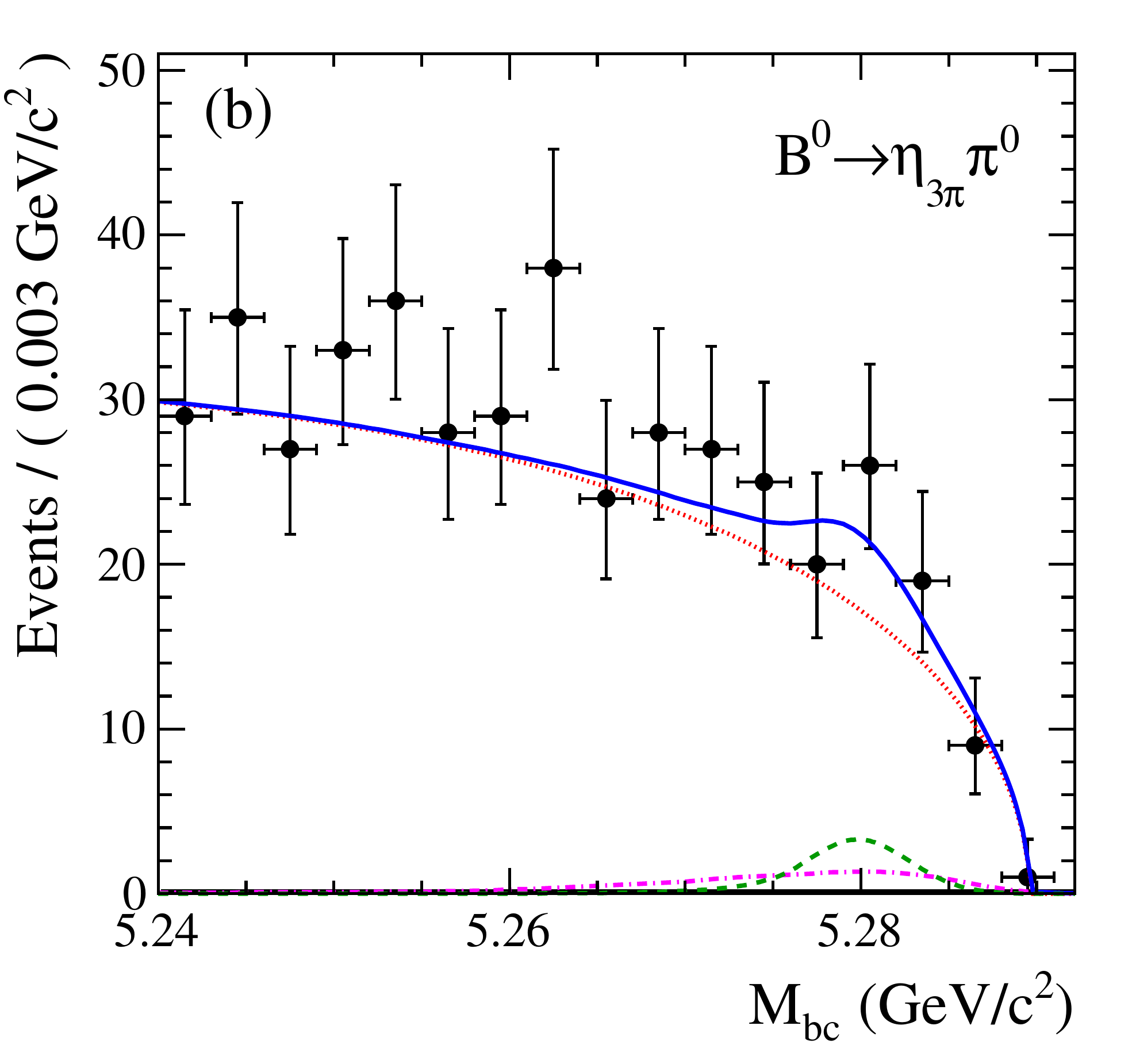}
    \includegraphics[width=0.35\textwidth]{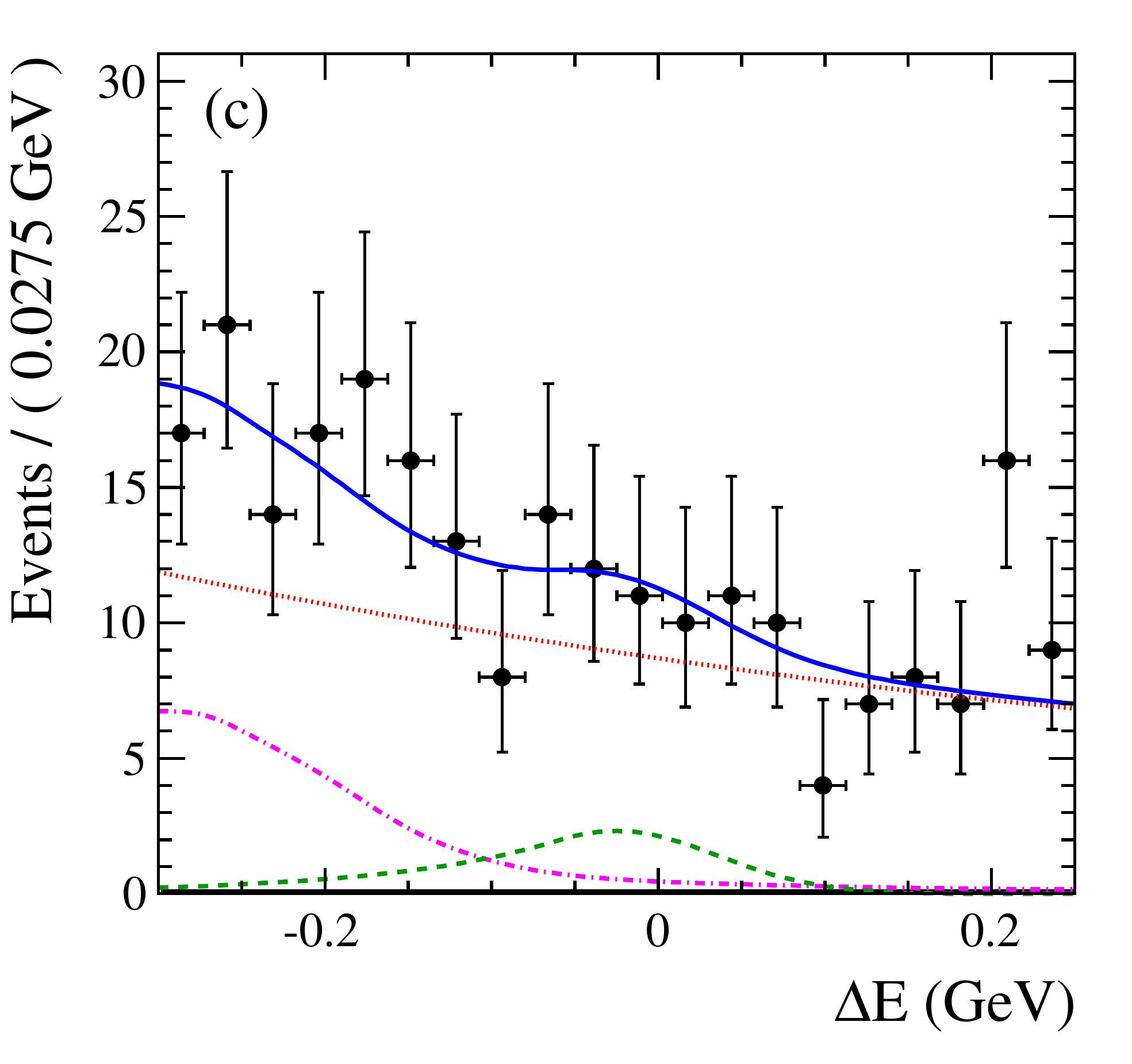}%
   \includegraphics[width=0.35\textwidth]{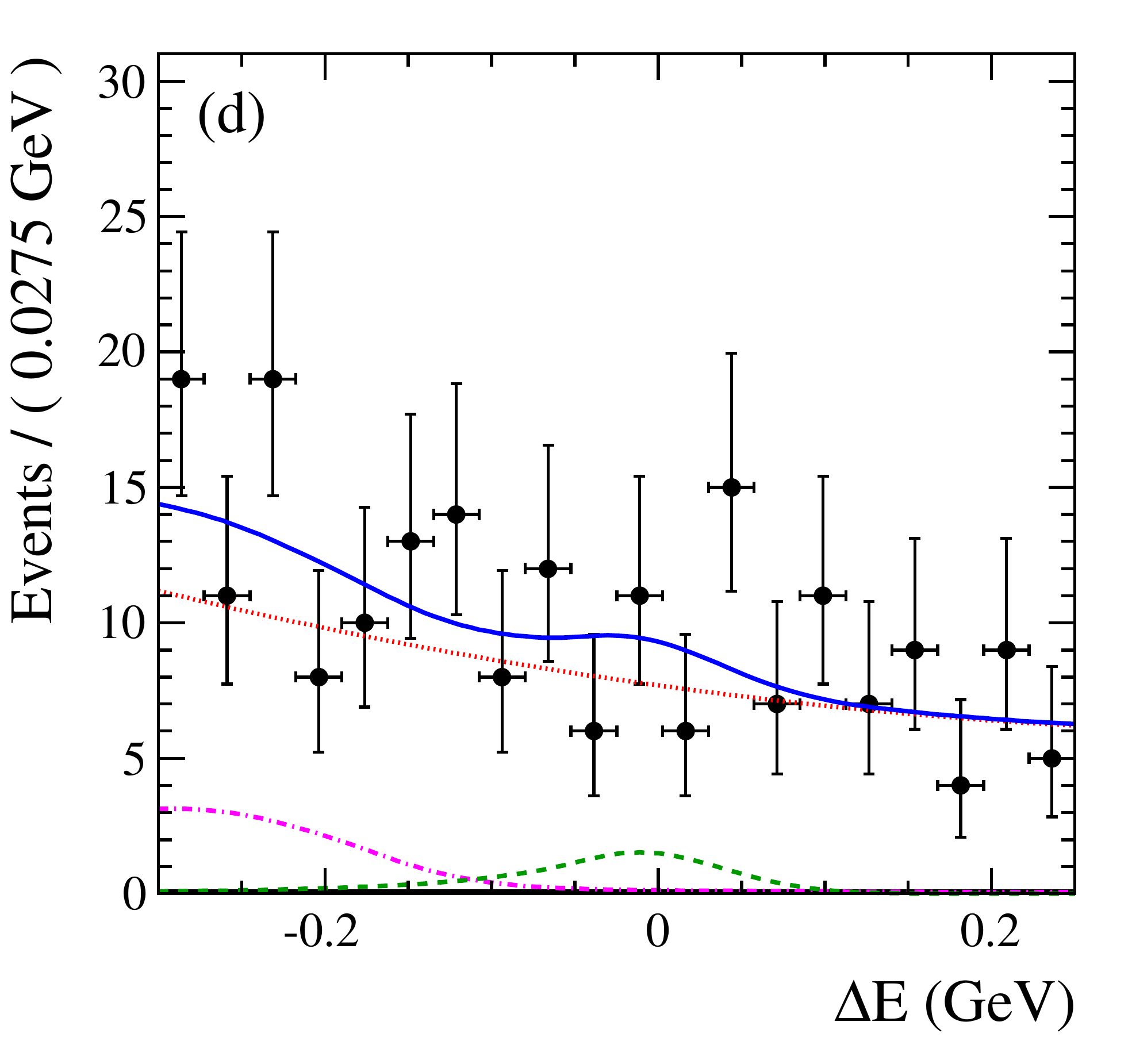}
    \includegraphics[width=0.35\textwidth]{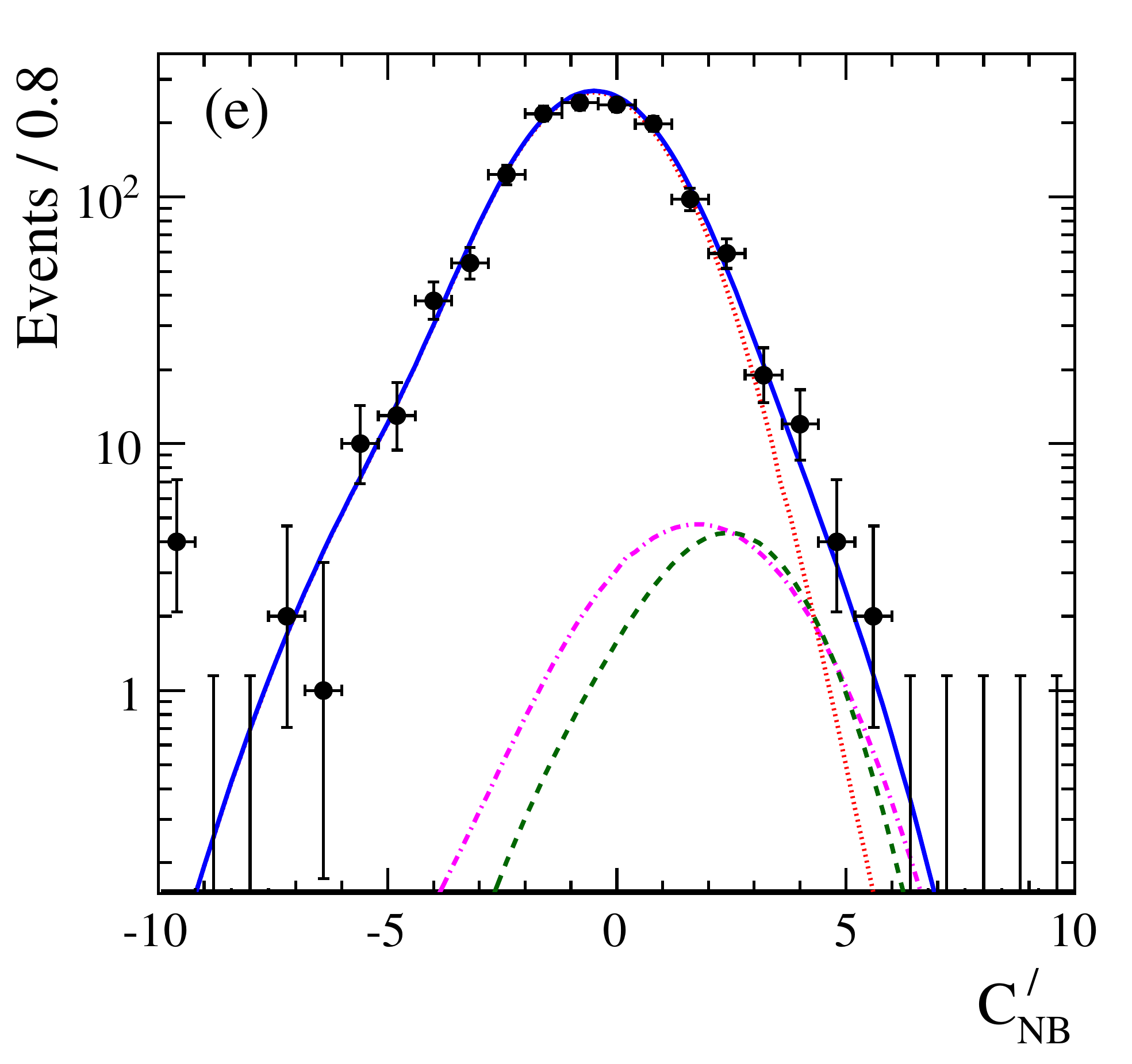}%
    \includegraphics[width=0.35\textwidth]{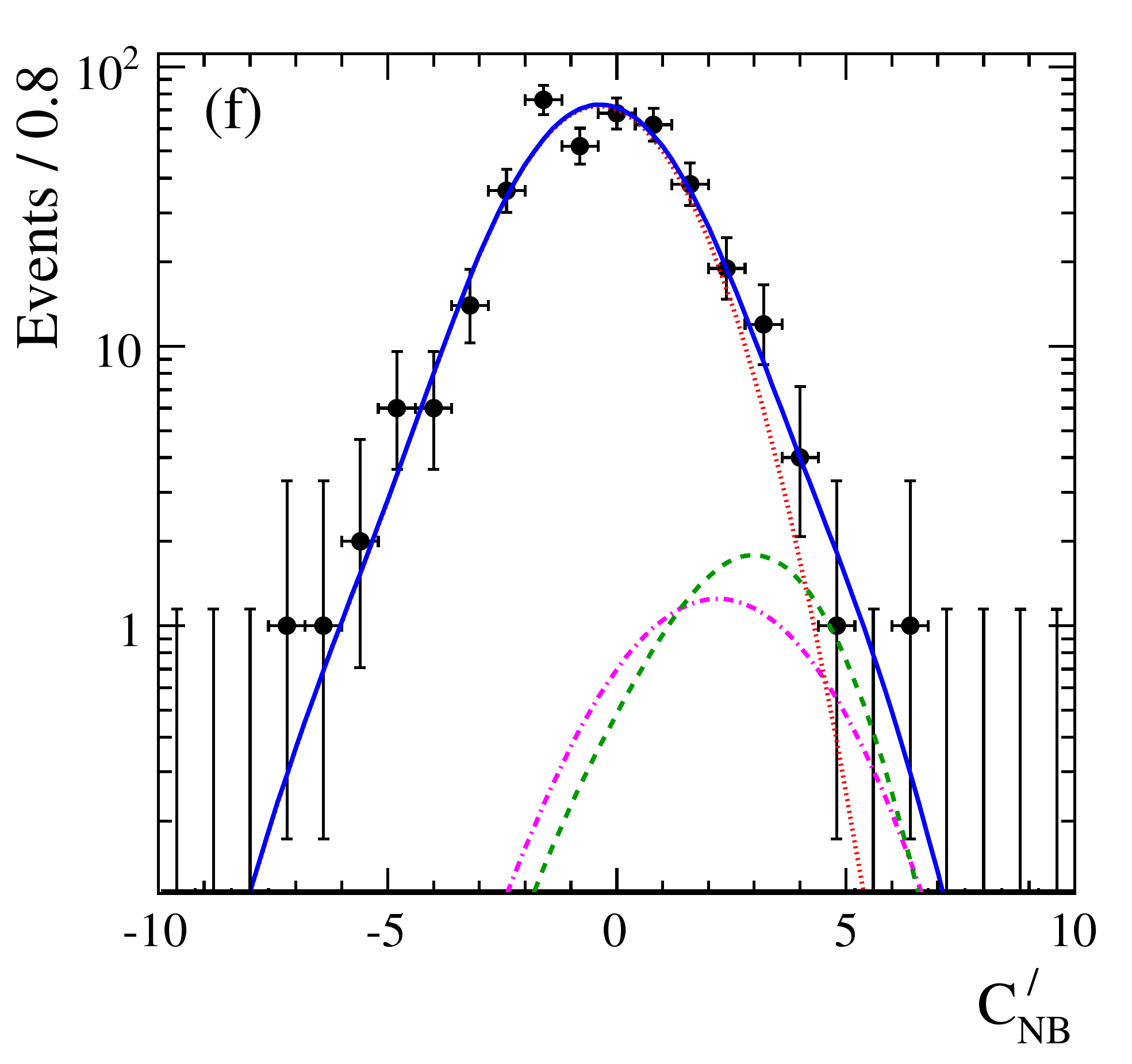}
\end{center}
\vskip -0.5cm
\caption{\small Projections of the simultaneous fit:
(a), (b) $M_{\rm bc}$; (c), (d) $\Delta E$;
(e), (f) $C'_{NB}$. Events plotted in $M_{\rm bc}$ $(\Delta E)$ 
are required to be in the signal region of $\Delta E$ $(M_{\rm bc})$, and  $C'_{NB}>1.5~ (0.5)$ for $\eta_{\gamma\gamma}\pi^0$  $(\eta_{3\pi}\pi^0)$ decays.  $C'_{NB}$ is plotted in the signal region of $M_{\rm bc}$ and  $\Delta E$.
The left (right) column corresponds to $\eta\to\gamma\gamma$
($\eta\to\pi^+ \pi^-\pi^0$) decays. 
Points with error bars are data;  the (green) dashed, (red) dotted and  (magenta) dot-dashed  curves represent the signal,  continuum and charmless rare  backgrounds, respectively, and the (blue) solid curves represent
the total PDF.}
\label{fig:real_full}
\end{figure}

The signal yields obtained from the fits are listed in
Table~\ref{fit}. The resulting branching fractions are
calculated as
\begin{linenomath}
\begin{equation}
\mathcal{B}(B^0\to\eta\pi^0) = \frac{Y_{\rm sig}}
{N_{B\Bbar}\times \epsilon\times\mathcal{B}_{\eta}},\label{eq:br}
\end{equation}
\end{linenomath}
where
$Y_{\rm sig}$ is the fitted signal yield,
$N_{B\Bbar} = (753\pm10)\times10^6$  is the number of $B\Bbar$ events,
$\epsilon$ is the signal efficiency as obtained from MC simulations,
and $\mathcal{B}_{\eta}$ is the branching fraction for $\eta\to\gamma\gamma$
or $\eta\to\pi^+\pi^-\pi^0$~\cite{PDG}. For the latter mode,
$\epsilon$ is corrected by a factor $\epsilon_{\rm PID} = 0.955 \pm 0.015$
to account for a small difference in particle identification (PID) 
efficiencies between data and simulations. This correction is estimated from
a sample of $D^{*+}\to D^0(\to K^-\pi^+)\pi^+$ decays.
In Eq.~(\ref{eq:br}), we assume equal production of $B^0\Bbar^0$ and $B^+B^-$
pairs at the $\Upsilon(4S)$ resonance. The combined branching fraction
is determined by simultaneously fitting both $B^0\to\eta_{\gamma\gamma}\pi^0$
and $B^0\to\eta_{3\pi}\pi^0$ samples for a common  $\mathcal{B}(B^0\to\eta\pi^0)$.  Projections of the simultaneous fit are
shown in Fig.~\ref{fig:real_full}.

\begin{table}[htb]
\renewcommand{\arraystretch}{1.5}
\caption{\small Fitted signal yield $Y_{\rm sig}$,
reconstruction efficiency $\epsilon$,
$\eta$ decay branching fraction $\mathcal{B}_{\eta}$,
signal significance, and $B^0$ branching fraction 
$\mathcal{B}$. The errors listed are statistical only.
The significance  includes both statistical and
systematic uncertainties (see text).}
\label{fit}
\begin{tabular}{c|ccccc}
\hline \hline
Mode & $Y_{\rm sig}$ & $\epsilon(\%)$ & $\mathcal{B}_{\eta}(\%)$ & 
Significance & $\mathcal{B}(10^{-7})$\\
\hline
$B^0\to\eta_{\gamma\gamma}\pi^0$ & $30.6^{+12.2}_{-10.8}$ &
       18.4 & 39.41 & 3.1 & $5.6^{+2.2}_{-2.0}$ \\
$B^0\to\eta_{3\pi}\pi^0$ & $0.5^{+6.6}_{-5.4}$ &
       14.2 & 22.92 & 0.1 & $0.2^{+2.8}_{-2.3}$ \\
Combined & & & & 3.0 & $4.1^{+1.7}_{-1.5}$ \\
\hline\hline
\end{tabular}
\end{table}

The signal significance is calculated as $\sqrt{-2\ln(\mathcal{L}_0/\mathcal{L}_{\rm max})}$,
where $\mathcal{L}_0$  is the likelihood value when the
signal yield is fixed to zero, and  $\mathcal{L}_{\rm max}$ is the likelihood 
value of the nominal fit.
To include systematic uncertainties in the significance,
we convolve the likelihood distribution with a Gaussian function
whose width is set  to the total systematic uncertainty
that affects  the signal yield. The resulting significance 
is 3.0 standard deviations; thus, our measurement
constitutes the first evidence for this decay mode.  
A  Bayesian upper limit on the branching fraction is obtained
by integrating the likelihood function from zero to infinity;
the value that corresponds to 90\% of this total area is taken
as the 90\% C.L. upper limit. 
The result is $\mathcal{B}(B^0\to\eta\pi^0)<6.5\times 10^{-7}$.


The systematic uncertainty on the branching fraction
has several contributions, as listed in Table~\ref{sys1}. 
The systematic uncertainty due to the fixed parameters in the PDF 
is estimated by  varying them individually  according  to their 
 statistical uncertainties.
The resulting changes in the branching fraction are added in quadrature and  the result is taken
as the systematic uncertainty. 
We evaluate in a similar manner the uncertainty due to
errors in the calibration factors. 
The sum in quadrature of these two contributions constitutes the 
uncertainty due to PDF parametrization.
We perform  large ensemble tests in order to verify the
stability of our fit model. We find a potential  bias of $-2.6\%$, which 
we attribute to our neglecting small correlations among the fitted observables.
We assign a 3\% systematic uncertainty for each reconstructed
$\eta\to\gamma\gamma$ or $\pi^0\to\gamma\gamma$ decay~\cite{Chang:2012gnb} . The systematic uncertainty due to the
track reconstruction efficiency is 0.35\% per track,
as determined from a study of partially reconstructed
$D^{*+}\to D^0(\to K^0_S\pi^+\pi^-)\pi^+$ decays. A 1.6\%
uncertainty (0.8\% per pion) is assigned due to the
PID criteria applied to charged
pions in $\eta\to\pi^+\pi^-\pi^0$ decays. 
We determine the systematic uncertainty due to the $C_{NB}$
 selection  by applying different $C_{NB}$ criteria
and comparing the results with that of the nominal selection. The 
 differences observed are assigned as the systematic uncertainty.
The uncertainty due to the number of $B\Bbar$ pairs is 1.3\%, and
the uncertainty on $\epsilon$ due to MC statistics is 0.4\%. The
total systematic uncertainty is obtained by summing in quadrature
all  individual contributions. 

\begin{table}[htb]
\renewcommand{\arraystretch}{1.2}
\caption{\small Systematic uncertainties on $\mathcal{B}(B^0\to\eta\pi^0)$.
Those listed in the upper section are associated with fitting for the 
signal yields and are included in the signal significance and upper limit calculation. }
\label{sys1}
\centering
\begin{tabular}{l|c}
\hline \hline
Source & Uncertainty (\%) \\
\hline
PDF parametrization & $^{+10.2}_{-~9.2}$\\
Fit bias            & $^{+0.0}_{-2.6}$\\
\hline
$\pi^0 / \eta\to\gamma\gamma$ reconstruction & 6.0 \\
Tracking efficiency & 0.3 \\
PID efficiency & 0.6\\
$C_{NB}$ selection efficiency &$^{+2.1}_{-2.2}$ \\ 
MC statistics & 0.4 \\
Nonresonant contributions &$^{+~0.0}_{-10.8}$ \\
$\mathcal{B}(\eta\to\gamma\gamma)$ & 0.5 \\
$\mathcal{B}(\eta\to\pi^+\pi^-\pi^0)$ & 1.2 \\
Number of $B\Bbar$ pairs & 1.3 \\
\hline
Total & $^{+12.2}_{-15.9}$\\
\hline\hline
\end{tabular}
\end{table}

In order to check the reliability of the PDFs used for backgrounds,
we fit the data in the $M_{\rm bc}$ sideband 5.24-5.26 ${\rm GeV}/c^{2}$, where
 the end point of the ARGUS function used for the
continuum $M_{\rm bc}$ PDF is allowed to float. For all  three distributions, $M_{\rm bc}$, $\Delta E$, and $C'_{NB}$,
the MC PDFs give an excellent description of the data. We
subsequently fit a sample of MC sideband events constructed with
the same admixture of backgrounds as found in the data
sideband and obtain signal yields consistent with zero.

\begin{figure}[h!t!p!]
\begin{center}
    \includegraphics[width=0.35\textwidth]{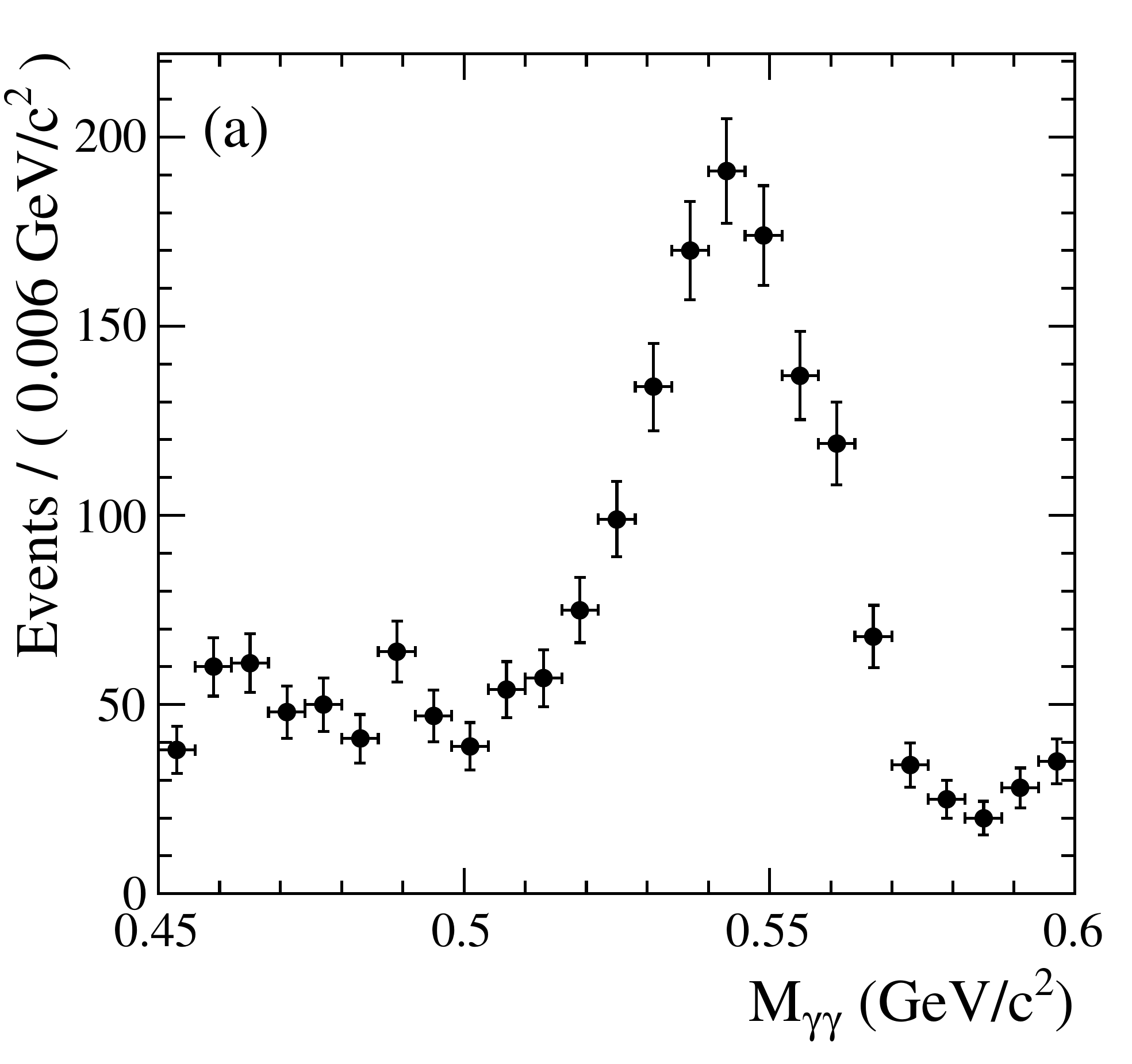}%
     \includegraphics[width=0.35\textwidth]{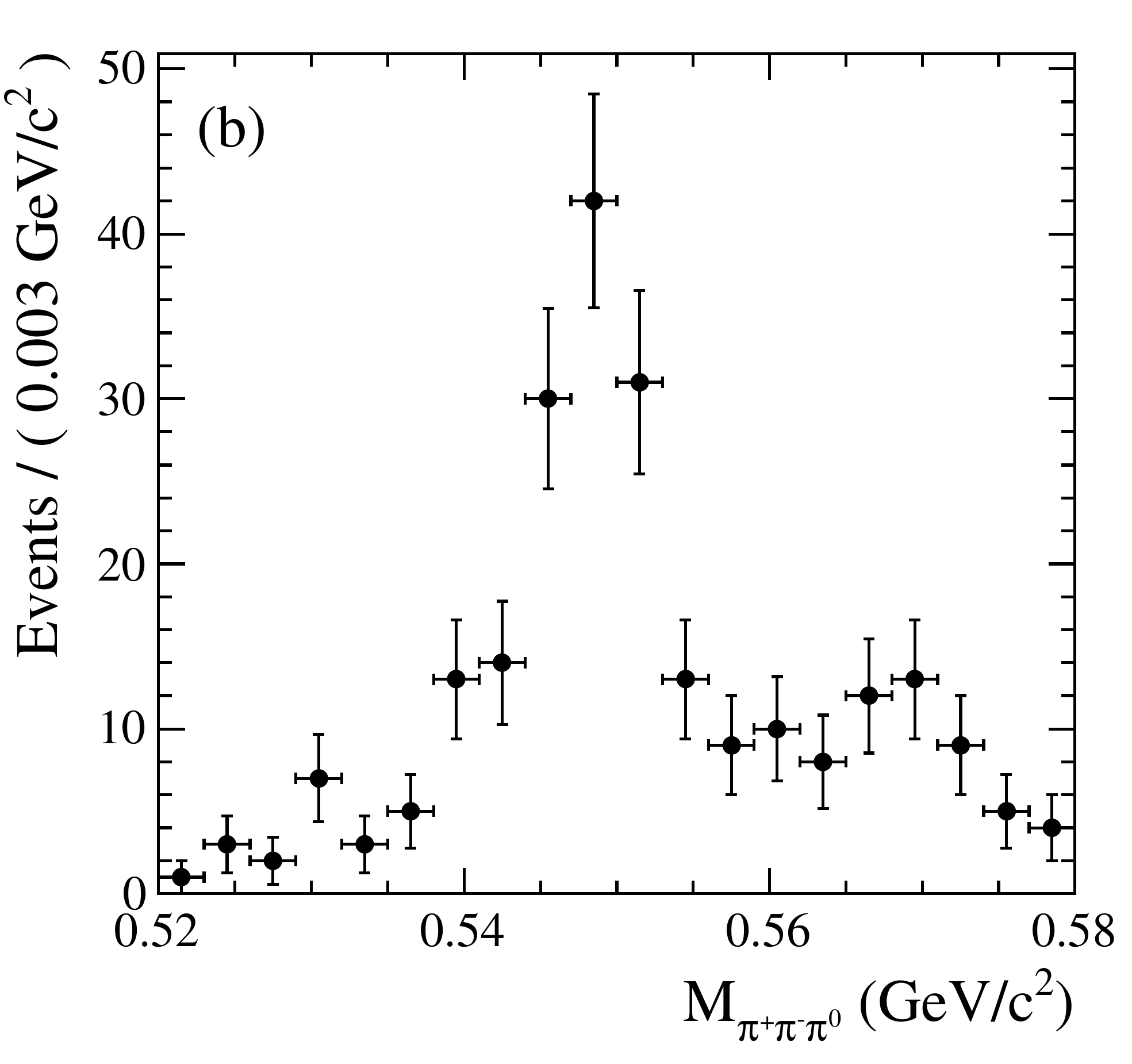}
\end{center}
\vskip -0.5cm
\caption{\small 
Distributions of (a) $M^{}_{\gamma\gamma}$ and 
(b) $M^{}_{\pi^+\pi^-\pi^0}$  invariant masses 
for events passing all selection requirements,
except those for $M^{}_{\gamma\gamma}$ or
$M^{}_{\pi^+\pi^-\pi^0}$.}
\label{fig:etamass}
\end{figure}

To check for potential nonresonant $B^0\to\gamma\gamma\pi^0$ 
and $B^0\to\pi^+\pi^-\pi^0\pi^0$ decays, we relax the 
$\eta$ mass requirement and 
plot the $\gamma\gamma$
and $\pi^+\pi^-\pi^0$ invariant mass distributions 
(Fig.~\ref{fig:etamass}) for events in the $M_{\rm bc}$--$\Delta E$ signal region. Significant peaks are
observed for $M^{}_{\gamma\gamma}\approx M^{}_\eta$
and $M^{}_{\pi^+\pi^-\pi^0}\approx M^{}_\eta$, as
expected. The small sidebands indicate no
significant contributions from nonresonant decays. 
We check this quantitatively 
by requiring that $M_{\gamma\gamma}$ $(M_{\pi^+\pi^-\pi^0})$ be
in the sideband  0.45-0.50 ${\rm GeV}/c^2$ (0.56-0.58 ${\rm GeV}/c^2$)
and repeat the fitting procedure. We find $2.2^{+4.8}_{-3.3}$ ($-2.2^{+3.4}_{-2.5}$) signal decays,
 consistent with zero. To be conservative, we assign  a systematic uncertainty  due to $B^0\to\gamma\gamma\pi^0$   nonresonant decays by appropriately scaling this fit result.

In summary, we report a measurement of the branching
fraction for $\bzetapiz$ decays. We obtain
\begin{eqnarray}
\mathcal{B}(B^0\to\eta\pi^0) & = & 
\left( 4.1^{+1.7+0.5}_{-1.5-0.7}\right) \times 10^{-7}\nonumber,
\end{eqnarray}
where the first uncertainty is statistical and the second
is systematic. This corresponds to a 90\% C.L. upper limit of $\mathcal{B}(B^0\to\eta\pi^0)<6.5\times 10^{-7}$. The significance of this result is~$3.0$ standard deviations,
and thus this measurement constitutes the first evidence for
this decay. The measured branching fraction  is in good agreement with theoretical
expectations~\cite{qcd, Williamson:2006hb, su3}. Inserting our measured value
into Eq. (19) of Ref.~\cite{Gronau:2005pq} gives the result that the
isospin-breaking correction  to the  weak phase $\phi_2$ measured in $B\to\pi\pi$ decays due to $\pi^0$--$\eta$--$\eta'$ mixing is less than $0.97^{\circ}$ at 90\% C.L.

\begin{center}
\textbf{ACKNOWLEDGMENTS}
\end{center}

We thank the KEKB group for the excellent operation of the
accelerator; the KEK cryogenics group for the efficient
operation of the solenoid; and the KEK computer group,
the National Institute of Informatics, and the 
PNNL/EMSL computing group for valuable computing
and SINET4 network support.  We acknowledge support from
the Ministry of Education, Culture, Sports, Science, and
Technology (MEXT) of Japan, the Japan Society for the 
Promotion of Science (JSPS), and the Tau-Lepton Physics 
Research Center of Nagoya University; 
the Australian Research Council and the Australian 
Department of Industry, Innovation, Science and Research;
Austrian Science Fund under Grants No.~P 22742-N16 and No.~P 26794-N20;
the National Natural Science Foundation of China under Contracts 
No.~10575109, No.~10775142, No.~10875115, No.~11175187, and  No.~11475187; 
the Ministry of Education, Youth and Sports of the Czech
Republic under Contract No.~LG14034;
the Carl Zeiss Foundation, the Deutsche Forschungsgemeinschaft
and the VolkswagenStiftung;
the Department of Science and Technology of India; 
the Istituto Nazionale di Fisica Nucleare of Italy; 
National Research Foundation (NRF) of Korea Grants
No.~2011-0029457, No.~2012-0008143, No.~2012R1A1A2008330, 
No.~2013R1A1A3007772, No.~2014R1A2A2A01005286,   No.~2014R1A2A2A01002734,  and 
No.~2014R1A1A2006456;
the Basic Research Lab program under NRF Grants  No.~KRF-2011-0020333 and 
No.~KRF-2011-0021196, Center for Korean J-PARC Users, Grant  No.~NRF-2013K1A3A7A06056592; 
the Brain Korea 21-Plus program and the Global Science Experimental Data 
Hub Center of the Korea Institute of Science and Technology Information;
the Polish Ministry of Science and Higher Education and 
the National Science Center;
the Ministry of Education and Science of the Russian Federation and
the Russian Foundation for Basic Research;
the Slovenian Research Agency;
the Basque Foundation for Science (IKERBASQUE) and 
the Euskal Herriko Unibertsitatea (UPV/EHU) under program UFI 11/55 (Spain);
the Swiss National Science Foundation; the National Science Council
and the Ministry of Education of Taiwan; and the U.S.\
Department of Energy and the National Science Foundation.
This work is supported by a Grant-in-Aid from MEXT for 
Science Research in a Priority Area (``New Development of 
Flavor Physics'') and from JSPS for Creative Scientific 
Research (``Evolution of Tau-lepton Physics'').


\begin{thebibliography}{99}
\bibitem{charge-conjugate}
  Charge conjugate modes are implicitly included unless
  stated otherwise.
\end{thebibliography}
\end{document}